\documentclass[a4paper,11pt]{article}
\pdfoutput=1 

\usepackage{jcappub} 

\usepackage[T1]{fontenc} 

\usepackage{subfigure}
\usepackage{enumitem}
\usepackage{pifont}
\setlist[itemize]{itemsep=0.5ex,parsep=0pt,label=\checkmark}

\usepackage{color, colortbl}
\usepackage[dvipsnames]{xcolor}
\definecolor{Gray}{gray}{0.9}
\definecolor{Gray2}{gray}{0.95}
\definecolor{LightCyan}{rgb}{0.88,1,1}

\usepackage[normalem]{ulem} 

\title{\boldmath Hunting the stochastic gravitational wave background \\ in pulsar timing array cross correlations through theoretical uncertainty}

\author[a,1]{Reginald Christian Bernardo,\note{Corresponding author.}}
\author[a,b]{and Kin-Wang Ng}

\affiliation[a]{Institute of Physics, Academia Sinica, \\ Taipei 11529, Taiwan}
\affiliation[b]{Institute of Astronomy and Astrophysics, Academia Sinica, \\ Taipei 11529, Taiwan}

\emailAdd{rbernardo@gate.sinica.edu.tw}
\emailAdd{nkw@phys.sinica.edu.tw}

\abstract{
Incredible progress on the theoretical uncertainty of the spatial correlations of the stochastic gravitational wave (GW) background were recently made. However, it remains to realize the impact of this theoretical uncertainty on PTA cross correlations analysis. This paper pushes forward in this direction, as a proof--of--principle: showing the potential role that theoretical uncertainty has on unburying the stochastic GW background signal in noisy PTA cross correlation measurements. We consider both a mock data set and the noise--marginalized 12.5 years NANOGrav spatial correlation measurements, and find optimistic conclusions regardless of the physical content of the GW background and the nature of the noise in the data. Very briefly, we show through various cases a modest result that looking out for a stochastic signal is better when two of its moments are utilized. Or, in terms of GWs, we show that the theoretical uncertainty can play a substantial role in the hunt for the stochastic GW background.
}

\begin{document}
\maketitle
\flushbottom

\section{Introduction}
\label{sec:introduction}

Gravity shaped the Universe and is arguably the most fundamental of the forces of nature. However, our best description of it based on general relativity (GR) can only bring us so close to a fundamental understanding of gravity in terms of particle, or string context or however it should be perceived in quantum gravity. Gravitational waves (GW) -- spacetime fluctuations, or gravitational radiation propagating through spacetime itself, emitted by masses, or gravitational charges in acceleration -- could just be the best probe for looking into the basic properties of gravity \cite{Cutler:2002me, Flanagan:2005yc, Barack:2018yly}, e.g., one subkilohertz-GW observation with an optical counterpart single-handedly constrained the GW speed to unmatched precision \cite{LIGOScientific:2017vwq}, significantly impacting the space of viable alternative gravity theories \cite{Lombriser:2015sxa, Sakstein:2017xjx, Creminelli:2017sry, Baker:2017hug, Ezquiaga:2017ekz}. Space-based millihertz GW detectors \cite{LISA:2017pwj, TianQin:2015yph} are expected to further complement the ground-based ones, forming an assembly of multiband GW laser interferrometer detectors, to further our understanding of gravity through a broader scope of astrophysical sources emitting GWs \cite{Wyithe:2002ep, Baker:2019nia, Barausse:2020rsu}.

On the other hand, in the much lower nanohertz GW band \cite{2021NatRP...3..344B}, the multiband picture is drawn by pulsar timing arrays (PTA) \cite{NANOGrav:2020bcs, Goncharov:2021oub, Chen:2021rqp, 2010CQGra..27h4013H}, that utilize millisecond pulsars' timing residuals sensitivity to nanohertz GWs. Moreover, in this regime, the GWs are expected to superpose into a galactic size GW, so-called stochastic gravitational wave background (SGWB) \cite{Romano:2019yrj, Burke-Spolaor:2018bvk, Book:2010pf}, that encompasses the PTA and as a consequence spatially correlates the timing residuals between each pair of millisecond pulsars, i.e., SGWB produces a distinct two point correlation function. In GR, this correlation is embodied by the Hellings-Downs (HD) curve \cite{Jenet:2014bea, Romano:2016dpx}, and only the detection of this spatial correlation in PTA data would be the indispensable evidence for the existence of the SGWB \cite{Allen:2023kib}, which teases to be on the horizon of the PTAs sensitivity threshold \cite{NANOGrav:2020spf}. By and large, the observation of the HD can be regarded as the `holy grail' of PTA GW science.

The trail toward a detection of the SGWB is however far from straightforward due to the variety of astrophysical noise involved, among other factors such as clock and ephemeris uncertainty related ones \cite{2016MNRAS.455.4339T, Roebber:2019gha, NANOGrav:2020tig}. Nonetheless, an abundance of recent progress in the theoretical side were made, which first and foremost gives meaning to the data, and enriches the science that is learned with PTA. Without any claim to be an exhaustive list, we write below milestones we view have significantly advanced our theoretical understanding SGWB correlations:
\begin{itemize}
    \item 1979--Pulsar timing was proposed for the detection of nanohertz GWs \cite{Detweiler:1979wn};
    \item 1983--The SGWB correlation was derived \cite{Hellings:1983fr}, and have since been known as the HD curve;
    \item 2001--The spectral profiles were derived for SGWB given their sources \cite{Phinney:2001di};
    \item 2011--SGWB correlations were derived for non-Einsteinian GW polarizations propagating at the speed of light \cite{Chamberlin:2011ev};
    \item {2014--The power spectrum form of the HD correlation was derived \cite{Gair:2014rwa};}
    \item 2018--A power spectrum approach (PSA) was advanced for the calculation of SGWB correlations for non-Einsteinian GW modes \cite{Qin:2018yhy};
    \item 2020--PSA was generalized for subluminal SGWB correlations \cite{Qin:2020hfy};
    \item 2021--SGWB correlations were calculated for the massive gravity degrees of freedom \cite{Liang:2021bct};
    \item 2021--The PSA for luminal tensor GW modes was revisited, and generalized to finite pulsar distances \cite{Ng:2021waj};
    \item 2022--The variance of HD was calculated \cite{Allen:2022dzg};
    \item 2022--HD variance was generalized to consider arbitrary pulsar sky distributions \cite{Allen:2022ksj};
    \item 2022--The PSA was generalized for subluminal SGWB correlations by non-Einsteinian modes and finite pulsar distances \cite{Bernardo:2022rif};
    \item 2022--Variance of non-Einsteinian subluminal SGWB correlations were derived by the PSA \cite{Bernardo:2022xzl};
    \item {2023--Astronomical milestone detection by the PTAs of the SGWB \cite{NANOGrav:2023gor,Xu:2023wog,Antoniadis:2023lym,Reardon:2023gzh}.}
\end{itemize}

Clearly, while the above is incomplete, we see an appreciable amount of theoretical progress recently have been made, e.g., see also \cite{Siemens:2006yp, Guzzetti:2016mkm, Caprini:2018mtu, Tahara:2020fmn, Adshead:2021hnm, Garcia-Saenz:2022tzu, Chu:2021krj, Tasinato:2022xyq, Liu:2022skj, Bernardo:2022vlj, Liang:2023ary}, and that remains for PTA data to resolve. In this paper, focusing on tensor GWs superposing into the SGWB, we make use of the recent results on the variance of the correlations \cite{Allen:2022dzg, Bernardo:2022xzl} to demonstrate their potentially crucial part in the search for the SGWB. We view this as a reminder that a stochastic signal is made up of moments, and while the mean is often the most pronounced, the higher moments can support the searches for such a signal, and even give additional information about the nature of the fields involved.

The rest of this paper is structured as follows. We start with a concise discussion of the SGWB correlations and the recipe considered (in \href{https://ascl.net/2211.001}{PTAfast} \cite{2022ascl.soft11001B}) for the calculation of the relevant PTA observables (Section \ref{sec:sgwb_correlations}). By considering the HD curve together with mock (Section \ref{subsec:hdmock}) and actual PTA data (Section \ref{subsec:hd_nanograv12}), we then examine two ways the cosmic variance can be utilized for the SGWB searches through cross correlations. We then also look at the impact of the theoretical uncertainty on the statistical significance of subluminal GWs in the present data (Section \ref{sec:subluminal_gws}). We end with a summary of our output and tease future directions (Section \ref{sec:discussion}). {We encourage the interested reader to download our codes and analysis notebooks in the \href{https://github.com/reggiebernardo/PTAfast/tree/main/app2_cosmic_variance}{PTAfast GitHub repository}.}

\section{SGWB correlations}
\label{sec:sgwb_correlations}

We devote this brief section to an overview of SGWB phenomenology, utilizing the {power spectrum approach} \cite{Bernardo:2022rif, Bernardo:2022xzl}.

The SGWB in a PTA is expected to exhibit a characteristic time of arrival perturbation in the radio pulses emitted by {galactic} millisecond pulsars. This spatial correlation, $\gamma_{ab}\left(\zeta\right)$, can be characterized by its first two moments, the mean and cosmic variance \cite{Ng:1997ez}, given by
\begin{equation}
\label{eq:meancorr}
    \gamma_{ab}\left(\zeta\right) = \sum_{l} \dfrac{2l+1}{4\pi} C_l P_l\left(\cos \zeta\right) \,
\end{equation}
and
\begin{equation}
\label{eq:cosmicvariance}
    \Delta \gamma_{ab}^{\rm CV}\left(\zeta\right) = \sqrt{ \sum_{l} \dfrac{2l+1}{8\pi^2} C_l^2 P_l\left(\cos \zeta\right)^2} \,,
\end{equation}
respectively, where $\zeta$ is the angular separation of pulsars $a$ and $b$, $P_l(x)$'s are Legendre polynomials, and $C_l$'s are the SGWB angular power spectrum multipoles.

For the Hellings-Downs curve, the $C_l$'s are given by
\begin{equation}
    C_l^{\rm HD} = \dfrac{8\pi^{3/2}}{(l - 1) l (l + 1) (l + 2)} \,
\end{equation}
for $l \geq 2$. In general, for subluminal tensor modes, the power spectrum multipoles, $C_l$'s, can be shown to be \cite{Bernardo:2022rif}
\begin{equation}
    C_l = \dfrac{ J_l \left( fD_a \right) J_l^* \left( fD_b \right) }{\sqrt{\pi}} \,,
\end{equation}
where the $J_l(y)$'s are
\begin{equation}
    J_l\left( y \right) = \sqrt{2} \pi i^l \sqrt{ \dfrac{(l + 2)!}{(l - 2)!} } \int_0^{2\pi y v} \dfrac{dx}{v} e^{ix/v} \dfrac{j_l\left( x \right)}{x^2} \,.
\end{equation}
Above, $j_l(x)$'s are the spherical Bessel functions, $f$ is the GW frequency, $v$ is the GW speed, and $D_i$'s are the distances to the pulsars. The HD correlation appears in the rather special infinite distance and luminal speed limit of this general expression.

{We mention that the frequency dependence of the correlation factors in through the effective distance $fD$ and becomes pronounced only at subdegree pulsar pair angular separations, $\zeta \lesssim {\cal O}(1^\circ)$. This finite distance effect manifests as tiny wiggles in the correlation, e.g., Figure 2 of Ref. \cite{Bernardo:2022xzl}, or alternatively as a sustained growth in the power spectrum at higher multipoles, e.g., Figure 3 of Ref. \cite{Ng:2021waj}. Resolving this physical effect however demands a level of sensitivity such that the experimental noise become subdominant to the cosmic variance and that the data caters correlation measurements of many subdegree separated pulsar pairs. For the meantime, the data sets are insensitive to this effect.}

The overlap reduction function (ORF), $\Gamma_{ab}\left(\zeta\right) = \gamma_{ab}\left(\zeta\right) \times 0.5/\gamma_{ab}^{\rm HD}\left(0\right)$, traditionally normalized so that $\Gamma_{ab}^{\rm HD}\left(0\right) = 0.5$, represent the SGWB correlation in a PTA.

For this work, we utilize the public code PTAfast \cite{2022ascl.soft11001B} which performs all these calculations, and outputs the ORF together with its theoretical uncertainty either as a pulsar variance or cosmic variance \cite{Allen:2022dzg, Bernardo:2022xzl}.

\section{Hellings-Downs curve}
\label{sec:hd}

We show in this section that when the theoretical uncertainties are accounted for, the statistical significance of the SGWB correlations in noisy data notably improves. We consider the HD curve to establish a basis for this assertion, and use it in conjunction with mock correlations data (Section \ref{subsec:hdmock}) and the NANOGrav 12.5 year data set (Section \ref{subsec:hd_nanograv12}).

\subsection{HD in mock data}
\label{subsec:hdmock}

We generate a SGWB data with a GW strain amplitude $A_{\rm gw} = 10^{-15}$ by drawing pulsar pair samples from a Gaussian random distribution in angular correlation space with the HD curve and its cosmic variance. This is equivalent to an evenly scattered distribution of a hundred pulsars in the sky that is correlated by the SGWB. We then bury the SGWB data in a variety of correlation noise, described in detail in the following.

We consider a non-stationary noise, which is added to the SGWB correlation signal at each angle separation between pairs of pulsars. First, a stationary noise is obtained from a Gaussian distribution, $\Delta \Gamma_{ab}^{\rm SN} \sim {\cal N} \left( 0, 0.3 \right)$, that is then weighted by a frequency dependent function, $f_{\rm ns}(f) = f^{\alpha}/\left(f^{\alpha} + \left(\beta/f\right)^{\alpha-1} + \epsilon\right) + A_{\rm qns}\left(f/f_{\rm ref}\right)^2$, that increases the noise amplitude at low frequencies, resulting in a non-stationary noise, $\Delta \Gamma_{ab}^{\rm NSN} \left(\zeta\right)$. We add a small constant value to the noise to ensure that it is always positive in addition to some white noise, $\xi\left(\zeta\right) \sim {\rm W}\left(0, 1\right)$, that is added to the SGWB and the non-stationary noise, and further take into account outliers to the data, which are generated by adding a larger random value to a small subset of the data points. We lastly also consider offsets, $\Delta \Gamma_{ab}^{\rm OFF} \sim {\cal N}\left(0, 0.1\right)$, to the average and the errors in the data. {We emphasize that the frequency $f$ in this context are not related to GWs but are simply frequency values at which a nonstationary noise is modulated and added to the mock GW correlations. It does not represent any physical source but rather serves as a parameter for the noise generation process.}

Overall, our mock correlations data for each pulsar pair constitutes
\begin{equation}
\begin{split}
    \Gamma_{ab}^{\rm mock}\left(\zeta\right) = \ &
    \Gamma_{ab}^{\rm SGWB}\left(\zeta\right) + r_1^2 \left( \Delta \Gamma_{ab}^{\rm NSN}\left(\zeta\right) + \Delta \Gamma_{ab}^{\rm OFF}\left(\zeta\right) \right) + r_2^2 \xi\left(\zeta\right) \,.
\end{split}
\end{equation}
We shall consider $r_1 = r_2 = 0.1$ for the noise amplitudes, and for the weighting function, $\alpha = 1.5$, $\beta = 2.5$, $A_{\rm qns} = 0.5$, and $f_{\rm ref} = 1 \ {\rm yr}^{-1}$, and take $N_{\rm plsr} = 100$ pulsars, or rather $N_{\rm plsr} (N_{\rm plsr} - 1)/2$ pulsar pairs. We stress that the choice of these noise parameters are inconsequential for the purpose of this work and that our conclusions hold regardless of their actual values.

We now search for the SGWB signal in the data, ${\rm CCP}\left(\zeta\right) = A_{\rm gw}^2 \Gamma_{ab}^{\rm mock} \left( \zeta \right)$, where $A_{\rm gw} = 10^{-15}$, by performing a Bayesian analysis given a correlations model, $\Gamma_{ab}\left(\zeta\right)$, i.e., the HD curve, that make up a simple one--dimensional parameter space, $A^2$, for the SGWB amplitude squared. We use the likelihood function
\begin{equation}
\label{eq:loglike}
    \log {\cal L} = -\dfrac{1}{2}\sum_{\zeta} \left(\dfrac{ A^2
 \Gamma_{ab}\left(\zeta\right) - {\rm CCP}\left(\zeta\right) }{ 
\Delta {\rm CCP}\left(\zeta\right) }\right)^2 \,
\end{equation}
to compare the model from the data, where the sum over $\zeta$ goes over the angular separations of all pulsar pairs. The parameter posterior $P(\theta|D)$ for parameter(s) $\theta$ is hereafter inferred by means of Bayes' theorem, $P\left(\theta|D\right) \propto P\left(\theta\right) {\cal L}$, by sampling over the parameter space $\{\theta_i\}$ with a prior $P\left(\theta\right)$. We make use of a Markov chain Monte Carlo (MCMC) algorithm \cite{Trotta:2008qt} with an uncompromising Gelman--Rubin convergence criterion $R - 1 = 5\times10^{-3}$, and take a flat prior $P\left(\theta = A^2\right) \in [0.01, 30] \times 10^{-30}$. We also vary the initial points in the region $A^2 \in (1, 2) \times 10^{-30}$ in order to warrant a more reliable convergence of the MCMC, and assess the models' goodness of fit using the reduced chi-squared measure, $\overline{\chi}^2 = \chi^2/n$, where $\chi^2 \sim -2 \log \hat{{\cal L}}$, $\hat{{\cal L}}$ is the best fit likelihood in the sample, and $n$ is the size of the data, e.g., a good fit has $\overline{\chi}^2 \lesssim 1$. {We also calculate the Bayesian evidences $Z$ \cite{Trotta:2008qt, Liddle:2007fy} to assess the models' statistical significance beyond pointwise insights given by $\overline{\chi}^2$.}

Now, the novelty we add in this correlations analysis is to directly take into consideration the cosmic variance, the second moment of the correlations, in the search for the SGWB in the data. We do so in two ways. In the first way, we replace the fixed HD curve by a Gaussian random distribution with the mean and the variance specified by the HD curve and its cosmic variance, that is, instead of the fixed curve, $\Gamma_{ab}\left(\zeta\right)$, that is traditionally used to search for the SGWB in PTA data, we consider a random Gaussian variable, $\Gamma_{ab}\left(\zeta\right) \rightarrow {\cal N} \left( \Gamma_{ab}\left(\zeta\right), \Delta \Gamma_{ab}^{\rm CV}\left(\zeta\right) \right)$, to represent the GW--induced correlations. In the second approach, which is perhaps the more canonical way to consider the theoretical uncertainty \cite{Allen:2023kib}, we take into account the cosmic variance as an uncertainty to the data, i.e., $\Delta {\rm CCP}\left(\zeta\right) + A_{\rm gw}^2 \Delta \Gamma_{ab}^{\rm CV}\left(\zeta\right)$, and then look for the SGWB, $A_{\rm gw}^2 \Gamma_{ab}\left(\zeta\right)$, in this data set with larger effective error bars. This second approach is also equivalent to just replacing the denominator, $\Delta {\rm CCP}\left(\zeta\right)$, in the likelihood \eqref{eq:loglike} by $\Delta {\rm CCP}\left(\zeta\right) + A_{\rm gw}^2 \Delta \Gamma_{ab}^{\rm CV}\left(\zeta\right)$.

\begin{figure}[h!]
    \centering
	\subfigure[ \ \ SGWB amplitude squared ]{
		\includegraphics[width = 0.5 \textwidth]{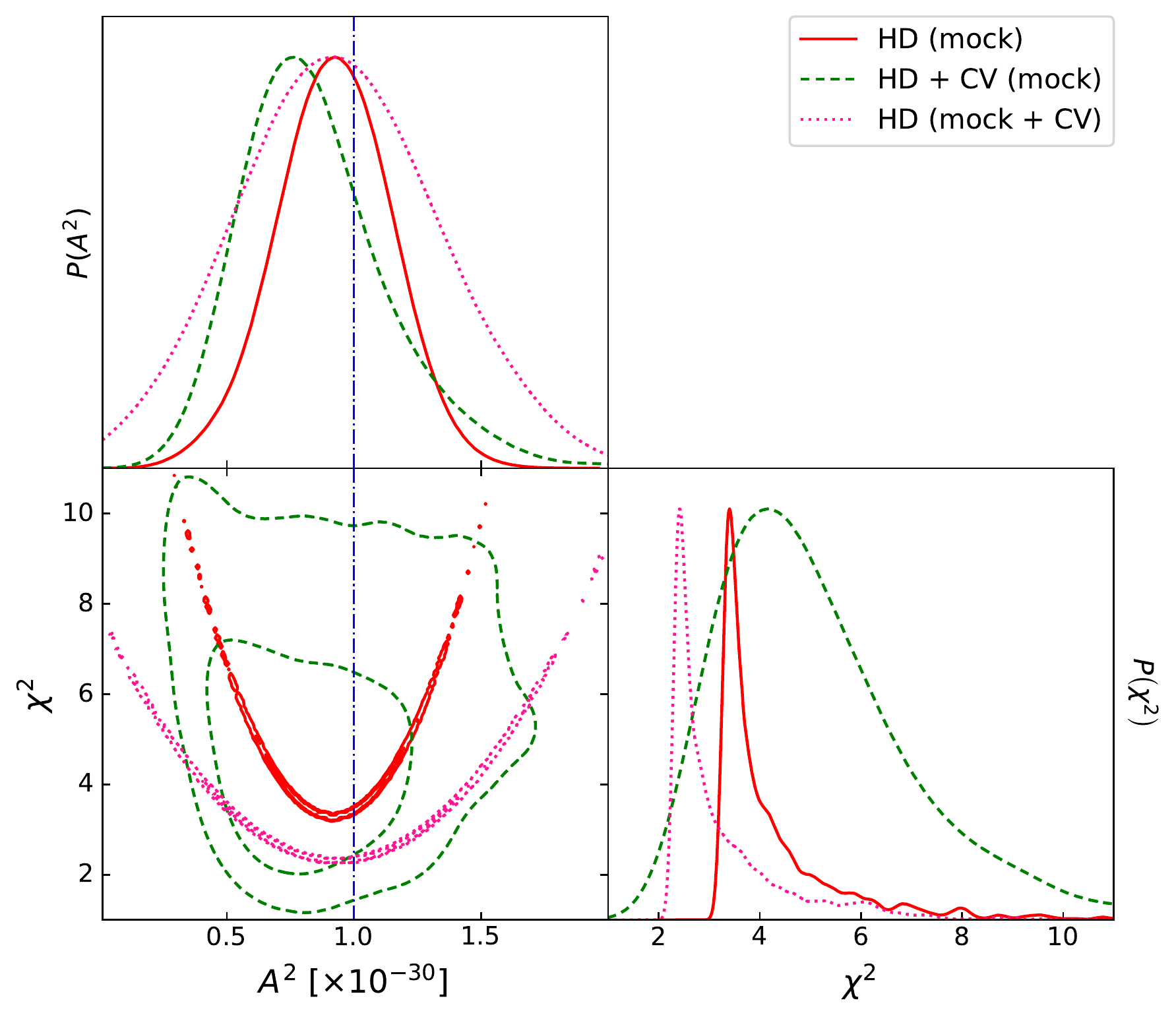}
		}
	\subfigure[ \ \ Best fit curves ]{
		\includegraphics[width = 0.45 \textwidth]{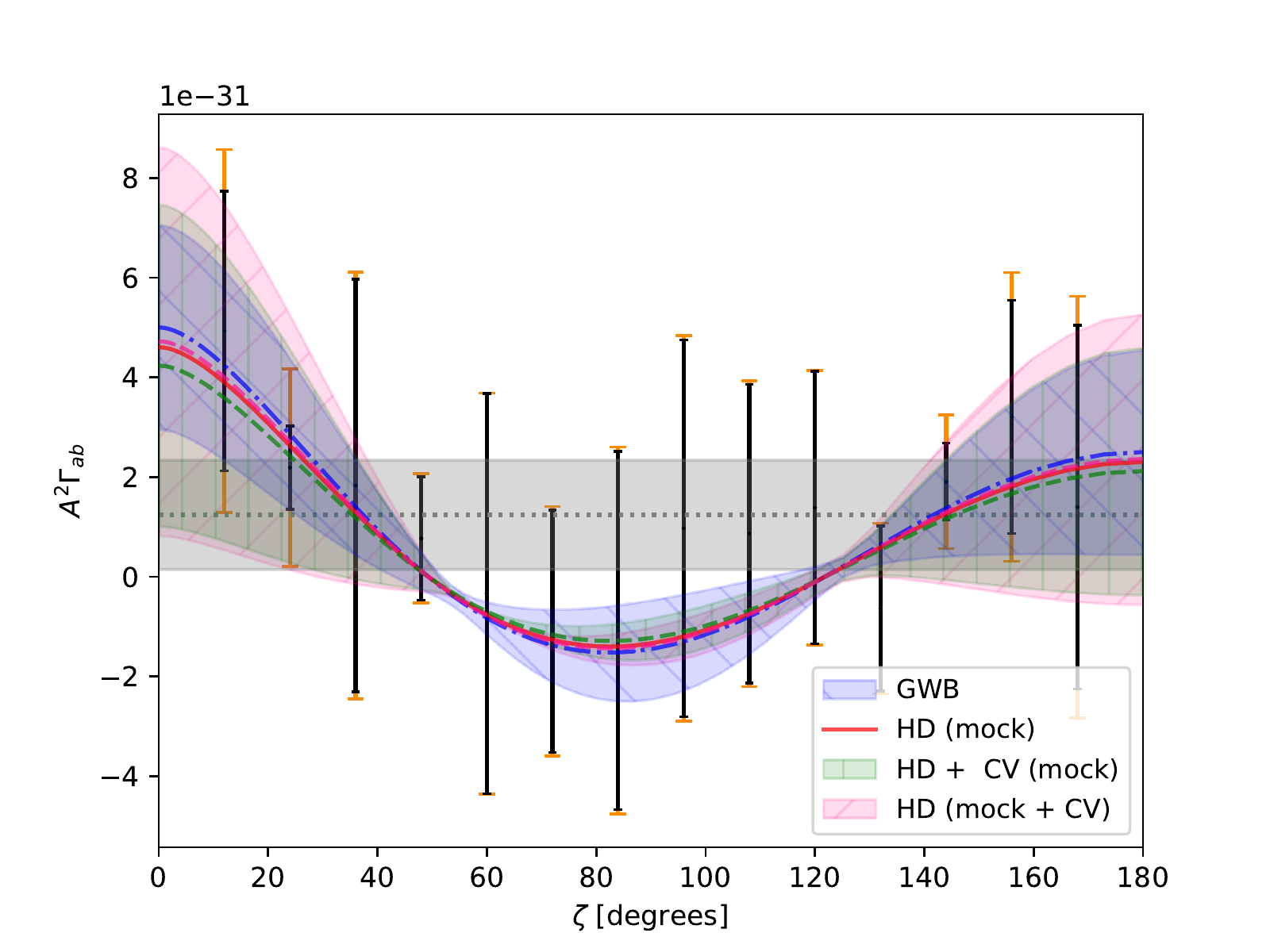}
		}
    \caption{Sampled posteriors and best fit curves for the HD correlation with mock correlations data (black bars, orange ones are effective error bars accounting for the cosmic variance). The true signal (GWB) is marked by the blue dash-dotted line, and best fit {uncorrelated Gaussian random noise (gray, dotted line shows mean, non-hatched horizontal band shows $1\sigma$)} is considered for reference. Colored--hatched bands show $1\sigma$ cosmic variance uncertainty from the mean of the correlation for the true signal (blue, `\textbackslash\textbackslash' hatches).}
    \label{fig:hd_mock_post}
\end{figure}

We find either ways of accounting for the theoretical uncertainty to be quite effective, as the results show in Figure \ref{fig:hd_mock_post} and Table \ref{tab:hdmockstats}. We denote by HD + CV (mock) the results coming from the first approach where the theoretical uncertainty is considered in the model, and by HD (mock + CV) the results from the second one where the theoretical uncertainty was taken together with the uncertainty in data.

It turns out that the inferred GW strain amplitudes are consistent within $68\%$ confidence limits with the true signal whether or not the cosmic variance is taken into account (Figure \ref{fig:hd_mock_post}(a)). The best fit curves in Figure \ref{fig:hd_mock_post}(b) support this and add that in fact the true signal is captured by both approaches, i.e., the best fit HD correlation curves with and without the consideration of the cosmic variance can be found in the bulk of the true signal. Of course this statement should be taken as a general observation rather than a strict rule, as obviously in certain places there are noticeable deviations. Nonetheless, the main point is clear that the parameter inferences and extracted SGWB signal is consistent with or without factoring in theoretical uncertainties. However, while this is true, it turns out that the cosmic variance can play a huge role in the search for the SGWB by improving the HD's statistical significance, as shown in Table \ref{tab:hdmockstats}.

\begin{table}[h!]
    \centering
    \caption{Marginalized statistics for HD cross correlations with mock data. The performance statistics {are the} reduced chi-squared, $\overline{\chi}^2 = \chi^2/n$ \cite{Trotta:2008qt, Liddle:2007fy}, {and the Bayesian evidence, expressed as $\log Z$, e.g., smaller $\overline{\chi}$ mean a better point estimate, and more positive $\log Z$ point to the better model. We additionally present the marginalized statistics for the monopole and an uncorrelated (Gaussian random) noise for comparison purposes}.}
    \begin{tabular}{|r|r|r|r|} \hline
    \phantom{$\dfrac{1}{1}$} model (data set) \phantom{$\dfrac{1}{1}$} & \phantom{$\dfrac{1}{1}$} $A^2$ [$\times 10^{-30}$] \phantom{$\dfrac{1}{1}$} & \phantom{$\dfrac{1}{1}$} {$\overline{\chi}^2$} \phantom{$\dfrac{1}{1}$} & \phantom{$\dfrac{1}{1}$} {$\log Z$} \phantom{$\dfrac{1}{1}$} \\ \hline \hline 
    \phantom{$\dfrac{1}{1}$} HD (mock) \phantom{$\dfrac{1}{1}$} & $1.01 \pm 0.20$ & {$0.23$} & {$-2.93 \pm 0.08$} \\ \hline
    \rowcolor{Gray}
    \phantom{$\dfrac{1}{1}$} \textbf{HD + CV (mock)} \phantom{$\dfrac{1}{1}$} & $\mathbf{0.85^{+0.21}_{-0.33}}$ & {$\mathbf{0.09}$} & {$\mathbf{-3.84 \pm 0.07}$} \\ \hline 
    \rowcolor{LightCyan}
    \phantom{$\dfrac{1}{1}$} \textbf{HD (mock + CV)} \phantom{$\dfrac{1}{1}$} & $\mathbf{0.94\pm0.39}$ & {$\mathbf{0.17}$} & {$\mathbf{-1.86 \pm 0.06}$} \\ \hline 
    \phantom{$\dfrac{1}{1}$} mon (mock) \phantom{$\dfrac{1}{1}$} & $0.19 \pm 0.07$ & {$0.49$} & {$-5.73 \pm 0.19$} \\ \hline 
    \phantom{$\dfrac{1}{1}$} {${\cal N}\left(\sigma < 0.42\right)$ (mock)} \phantom{$\dfrac{1}{1}$} & {$0.25 \pm 0.11$} & {$0.16$} & {$-7.41 \pm 0.07$} \\ \hline 
    \end{tabular}
    \label{tab:hdmockstats}
\end{table}

By now looking at the statistical significance, we are also able to distinguish between the two ways we accounted for the theoretical uncertainty. We see interestingly that the first approach, unstiffening the HD curve using the cosmic variance, give more statistical impact to the SGWB {in terms of the $\overline{\chi}^2$}. Nonetheless, we view the main result here is that regardless of how the theoretical uncertainty is taken into account, the {pointwise} statistical significance of the SGWB increases. {However, a slightly different picture comes out when the entirety of the sampled posteriors are considered  through the Bayesian evidence. In this way, we find that the unstiffening of the HD curve using the cosmic variances in fact weakens the evidence for the HD. We can understand this from Figure \ref{fig:hd_mock_post}(a) where despite having the best point estimate the broadening of the $\chi^2$ posterior has reduced the overall statistical evidence when the whole sampled space is taken into account. On the other hand, the evidence remains in favor of of the HD when the cosmic variance is canonically added to the noise in the data. As a matter of fact, treating the uncorrelated Gaussian random noise (${\cal N}(\sigma)$) as a null hypothesis improves the Bayes factor, $Z_{\rm HD}/Z_{\rm null}$, of the HD correlation from about $88$ to $257$ when the cosmic variance is considered\footnote{{The Bayes factor is calculated as $Z/Z_{\rm null}$ for a model with a Bayesian evidence $Z$ given a null hypothesis of evidence $Z_{\rm null}$ \cite{Trotta:2008qt, 2021JOSS....6.3001B}. The BF of the HD is then ${\rm BF}_{\rm HD} = \exp \left( \log Z_{\rm HD} - \log Z_{\rm null} \right) = \exp(-2.93 - (-7.41)) = 88.2$ without the CV, and ${\rm BF}_{\rm HD} = \exp(-1.86 - (-7.41)) = 257.2$ with the CV.}}. Granted, we find this in mock data, but as we are about to see the message is general.}

We understand that this exciting result is due to the theoretical uncertainty, represented in our case by the cosmic variance, being an extra layer of the stochastic signal, which in this case is the SGWB. Thus naturally searching for the SGWB in noisy correlations data becomes more efficient when both the first moment, the mean, and the second moment, the variance, of the correlation are given importance in the data analysis.

It should be noted that the monopole is irrelevant to this discussion, and only considered here for model comparison, which is reasonable since the flat--monopolar correlation signal is synonymous to nondetection of the SGWB. The result we point to here is that the goodness of fit (taken as a surrogate to statistical significance) of the HD correlation with the cosmic variance is better compared to the one without (Table \ref{tab:hdmockstats}). {We have also presented the statistical results for a Gaussian random noise which we take to depict an uncorrelated process.}

Clearly while we are making these statements with mock data sets with one kind of simulated noise, that is, a uniform one, all of the above remarks about the importance of the cosmic variance hold generally, as we proceed to show next, with genuine correlations data.

\subsection{HD in NG12}
\label{subsec:hd_nanograv12}

The leading PTAs observe with strong evidence a stochastic common spectrum process throughout the monitored millisecond pulsars \cite{NANOGrav:2020bcs, Chen:2021rqp, Goncharov:2021oub}. This common process can be speculated to originate from the SGWB; however, the more convincing observations unequivocally tied to SGWB are the spatial correlations shaped by the GWs based on their polarization content \cite{Allen:2023kib}. The challenge on this front is to dig up this stochastic--natured correlation from noise in the data that can be due to a variety of factors, understood or not. We show in this section that the consideration of the cosmic variance in the search of the SGWB with the NANOGrav 12.5 years correlations data set \cite{NANOGrav:2020bcs} improves the likelihood of the HD correlation. 

\begin{figure}[h!]
    \centering
	\subfigure[ \ \ SGWB amplitude squared ]{
		\includegraphics[width = 0.5 \textwidth]{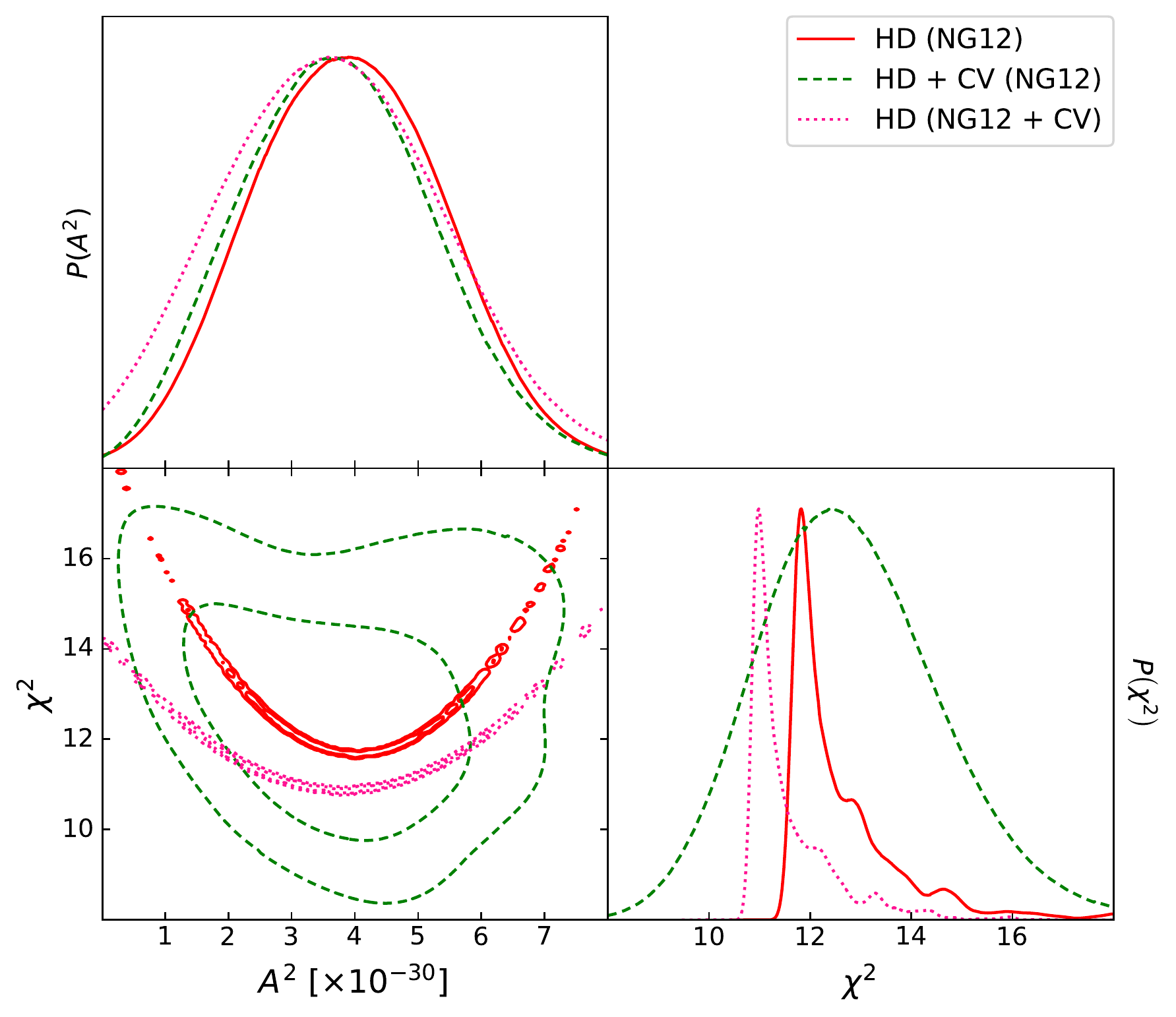}
		}
	\subfigure[ \ \ Best fit curves ]{
		\includegraphics[width = 0.45 \textwidth]{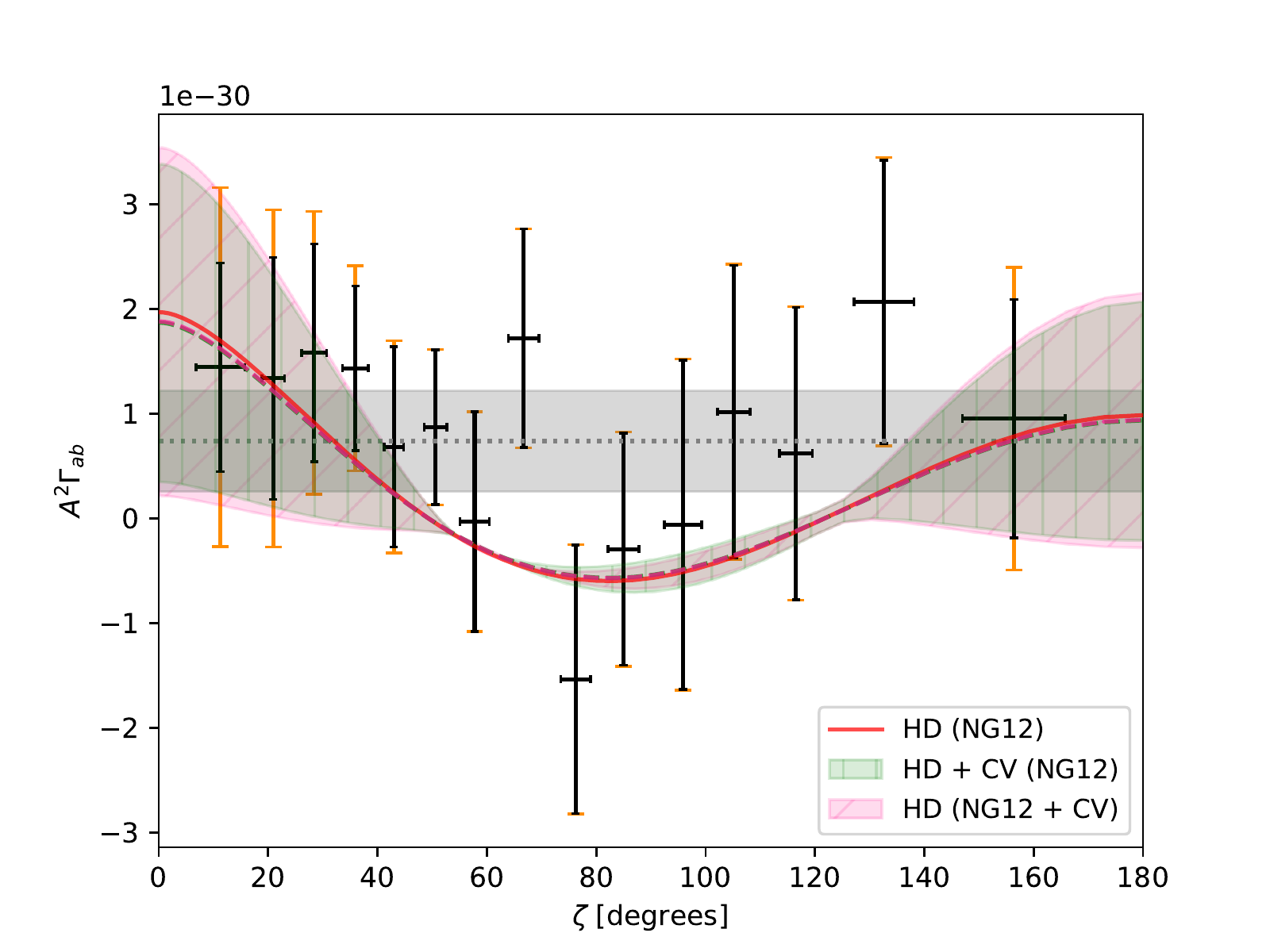}
		}
    \caption{Sampled posteriors and best fit curves for the HD correlation with the NANOGrav 12.5 years cross correlation measurements (black bars, orange ones are effective error bars accounting for the cosmic variance). The best fit {uncorrelated Gaussian random noise (gray, dotted line shows mean, non-hatched horizontal band shows $1\sigma$)} is shown for reference. Colored--hatched bands show $1\sigma$ cosmic variance uncertainty from the mean of the correlation.}
    \label{fig:hd_post}
\end{figure}

Proceeding as earlier, we perform a MCMC on the data set (NG12) with the models being the HD correlation without and with the cosmic variance. The sampled posteriors and best fit curves are shown in Figure \ref{fig:hd_post} and the marginalized statistics in Table \ref{tab:hdstats}. We denote by HD + CV (NG12) the results coming from the first approach where the correlation signal is given a random nature as $\Gamma_{ab}\left(\zeta\right) \sim {\cal N} \left(\Gamma_{ab}\left(\zeta\right), \Delta \Gamma_{ab}^{\rm CV}\left(\zeta\right)\right)$, and by HD (NG12 + CV) the results from the second approach where the theoretical uncertainty, $\Delta \Gamma_{ab}^{\rm CV}\left(\zeta\right)$, was considered with the uncertainty in the data.

We find this time broader posteriors for the GW amplitude squared which can be associated with the noise in the data. All the same, the estimates turn out consistent with or without the cosmic variance, as Figure \ref{fig:hd_post}(a) shows that the distributions nearly overlap. The best fit curves (Figure \ref{fig:hd_post}(b)) substantiate this further with the HD correlation best fit without the cosmic variance hovering about the mean of the corresponding best fit with the cosmic variance. Furthermore, both curves, HD + CV (NG12) and HD (NG12 + CV), accounting for the cosmic variance nearly just overlaps, with HD (NG12 + CV) only being slightly broader. So as far as the parameter estimates are concerned, the introduction of the cosmic variance to the correlations analysis do not bring too much new information, as we found with the previous mock data analysis. Alternatively, this can be taken to imply that the means of the correlation are sufficient to infer the relevant GW parameters from the cross correlation measurements.

\begin{table}[h!]
    \centering
    \caption{Best fit statistics for HD cross correlations constrained by PTA \cite{NANOGrav:2020bcs}. The performance statistics {are the} reduced chi-squared, $\overline{\chi}^2 = \chi^2/n$ \cite{Trotta:2008qt, Liddle:2007fy}, {and the Bayesian evidence, expressed as $\log Z$, e.g., smaller $\overline{\chi}$ mean a better point estimate, and more positive $\log Z$ point to the better model. We additionally present the marginalized statistics for the monopole and an uncorrelated (Gaussian random) noise for comparison purposes}.}
    \begin{tabular}{|r|r|r|r|} \hline
    \phantom{$\dfrac{1}{1}$} model (data set) \phantom{$\dfrac{1}{1}$} & \phantom{$\dfrac{1}{1}$} $A^2$ [$\times 10^{-30}$] \phantom{$\dfrac{1}{1}$} & \phantom{$\dfrac{1}{1}$} {$\overline{\chi}^2$} \phantom{$\dfrac{1}{1}$} & \phantom{$\dfrac{1}{1}$} {$\log Z$} \phantom{$\dfrac{1}{1}$} \\ \hline \hline 
    \phantom{$\dfrac{1}{1}$} HD (NG12) \phantom{$\dfrac{1}{1}$} & $3.9 \pm 1.5$ & {$0.78$} & {$-6.80 \pm 0.06$} \\ \hline
    \rowcolor{Gray}
    \phantom{$\dfrac{1}{1}$} \textbf{HD + CV (NG12)} \phantom{$\dfrac{1}{1}$} & $\mathbf{3.7 \pm 1.5}$ & {$\mathbf{0.61}$} & {$\mathbf{-7.23 \pm 0.09}$} \\ \hline
    \rowcolor{LightCyan}
    \textbf{HD (NG12 + CV)} \phantom{$\dfrac{1}{1}$} & $\mathbf{3.8^{+1.6}_{-2.0}}$ & {$\mathbf{0.73}$} & {$\mathbf{-6.12 \pm 0.07}$} \\ \hline 
    \phantom{$\dfrac{1}{1}$} mon \phantom{$\dfrac{1}{1}$} & $1.80 \pm 0.55$ & {$0.58$} & {$-6.40 \pm 0.10$} \\ \hline
    \phantom{$\dfrac{1}{1}$} {${\cal N}\left(\sigma < 0.34\right)$ (mock)} \phantom{$\dfrac{1}{1}$} & {$1.47 \pm 0.63$} & {$0.30$} & {$-8.47 \pm 0.41$} \\ \hline 
    \end{tabular}
    \label{tab:hdstats}
\end{table}

On the other hand, we find that the purpose of keeping the cosmic variance in the analysis is to boost the search of SGWB correlations by providing an accompanying information -- the variance of the signal -- that such a stochastic signal is expected to carry. This shows up in Table \ref{tab:hdstats} where clearly the HD correlation with or without theoretical uncertainty, as expected, appears to be a less favored model of the observed correlations than the monopole {and the Gaussian random noise, by the pointwise $\overline{\chi}^2$ standards}. However, we find that the statistical significance tilts to the favor of the HD when the cosmic variance is factored in the correlations model. In particular, the goodness of fit measures for the monopole and the randomized HD correlation with the cosmic variance, HD + CV (NG12), are comparable. The HD correlation with the cosmic variance (HD + CV, or ${\cal N} \left(\Gamma_{ab}^{\rm HD}\left(\zeta\right), \Delta \Gamma_{ab}^{\rm CV}\left(\zeta\right)\right)$) therefore is as compelling as the monopole a source of the observed spatial correlations. We also note that even with just considering the theoretical uncertainty as an effective error in the data, the significance of the HD improves. Again, regardless of the way we account for the theoretical uncertainty in the data analysis, we realize that it increases the {pointwise} significance of the stochastic signal that we take the SGWB to be.

{We look to the Bayesian evidence which sums up the significance over the entire sampled parameter space. In this view, we find the significance of the cosmic variance randomized HD to be lower compared to the HD without cosmic variance. We can understand this again by looking at Figure \ref{fig:hd_post}(a) where clearly the $\chi^2$ values corresponding to the unstiffened HD have become wider, thus explaining the higher pointwise likelihood but giving a lower significance over the whole sampled space. But then, the canonical accounting of the cosmic variance as an effective noise proves to be more promising in looking for the HD in noisy data. We find in this way that the Bayes factor of the HD correlation compared with the null hypothesis (uncorrelated Gaussain random noise) has increased from about $Z_{\rm HD}/Z_{\rm null} \sim 5-10$, which is admittedly not a lot, but it gives enough optimism about finding the HD with substantial evidence. Moreover, we see that the HD correlation with the cosmic variance as canonical noise is nearly as evident as the monopole, i.e., $Z_{\rm HD}/Z_{\rm mon} \sim 1$. This illustrates the role that the cosmic variance can play in finding the SGWB in noisy PTA correlations.}

In this section, we looked into the cosmic variance's influence to the HD correlation's performance as a model of observed, and mock, spatial correlations PTA data, and presented the optimistic results. We take this a step further in the following section by considering the impact of theoretical uncertainty to the correlations analysis with SGWB from subluminal GWs.

\section{Subluminal GW correlations}
\label{sec:subluminal_gws}

We test the robustness of the results of the previous section by checking out SGWB correlations beyond the HD that are allowed in the nanohertz GW regime \cite{Bernardo:2023mxc}. For this purpose, we consider subluminal GW modes with the NANOGrav 12.5 years spatial correlation measurements. The SGWB correlations model for this is described by the GW speed, $v$, modifying the shape of the ORF, $\Gamma_{ab}\left(\zeta\right)$, in addition to the GW strain amplitude squared that modulates the size of the correlations profile in pulsar angular separation space. Thus, this time, we perform an MCMC analysis over a two dimensional space, $A^2 \times v$, in contrast with the previously considered limiting case of the HD correlation ($v = 1$).

We take the theoretical uncertainty similarly as with the previous cases. In the first approach, we use the cosmic variance to soften the ORF, or randomizing it as $\Gamma_{ab}\left(\zeta\right) \rightarrow {\cal N}\left( \Gamma_{ab}\left(\zeta\right), \Delta \Gamma_{ab}^{\rm CV}\left(\zeta\right) \right)$, where now we emphasize that this ORF is also a function of the GW speed $v$. On the other hand, in the second approach, we consider the cosmic variance, $\Delta \Gamma_{ab}^{\rm CV}\left(\zeta\right)$, with the uncertainty of the data set. We present the results of both approaches taking into account the cosmic variance with the PTA data: T + CV (NG12) from the first approach, and T (NG12 + CV) from the second one.

In \cite{Bernardo:2023mxc}, we have shown that subluminal GW modes without the cosmic variance are disfavored by the NANOGrav 12.5 years noise--marginalized spatial correlation measurements. We shall see how this changes with the theoretical uncertainty factored in the correlations analysis.

\begin{figure}[h!]
    \centering
	\subfigure[ \ \ SGWB amplitude squared $\times$ GW speed ]{
		\includegraphics[width = 0.53 \textwidth]{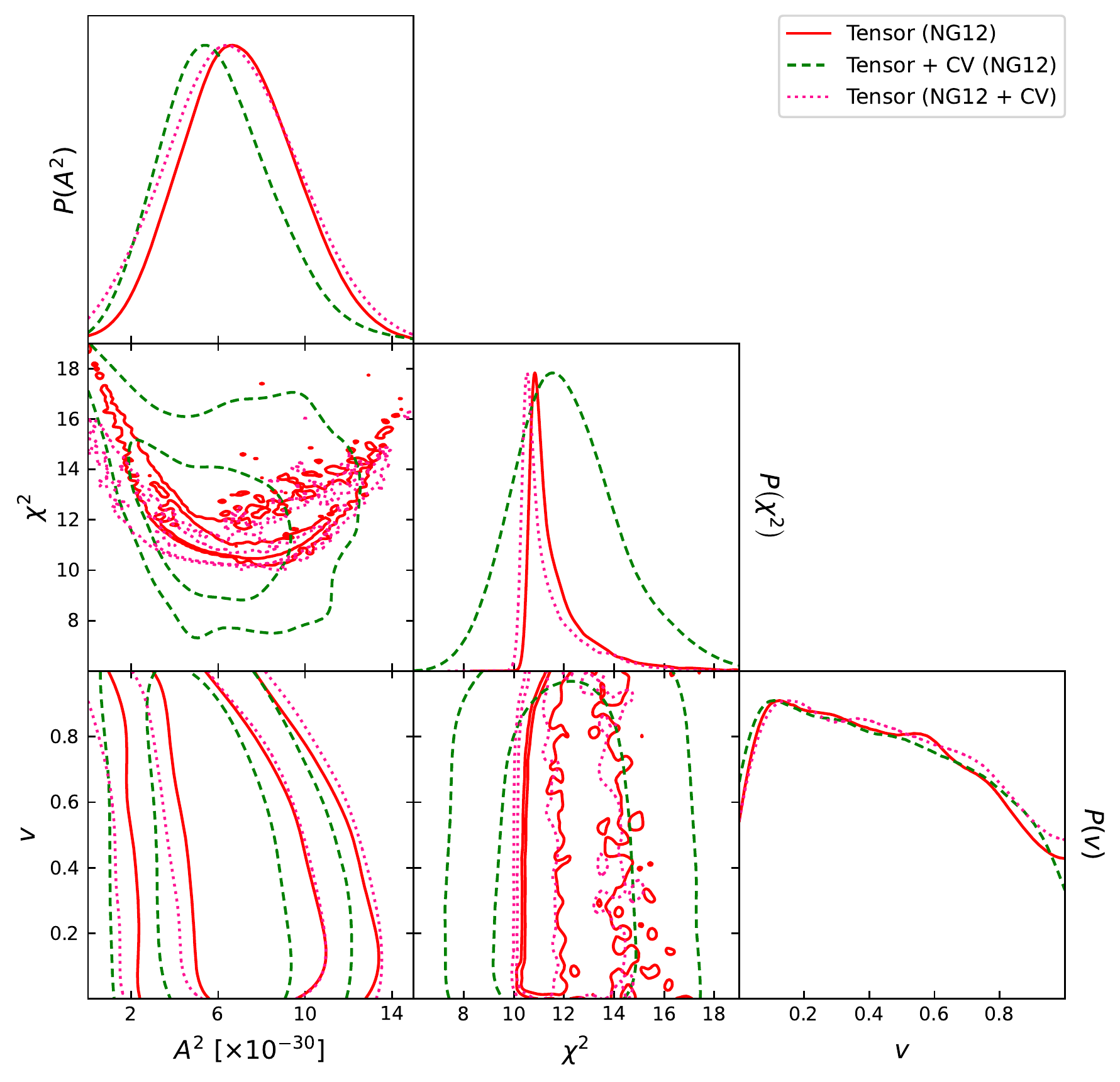}
		}
  	\subfigure[ \ \ Best fit curves ]{
		\includegraphics[width = 0.42 \textwidth]{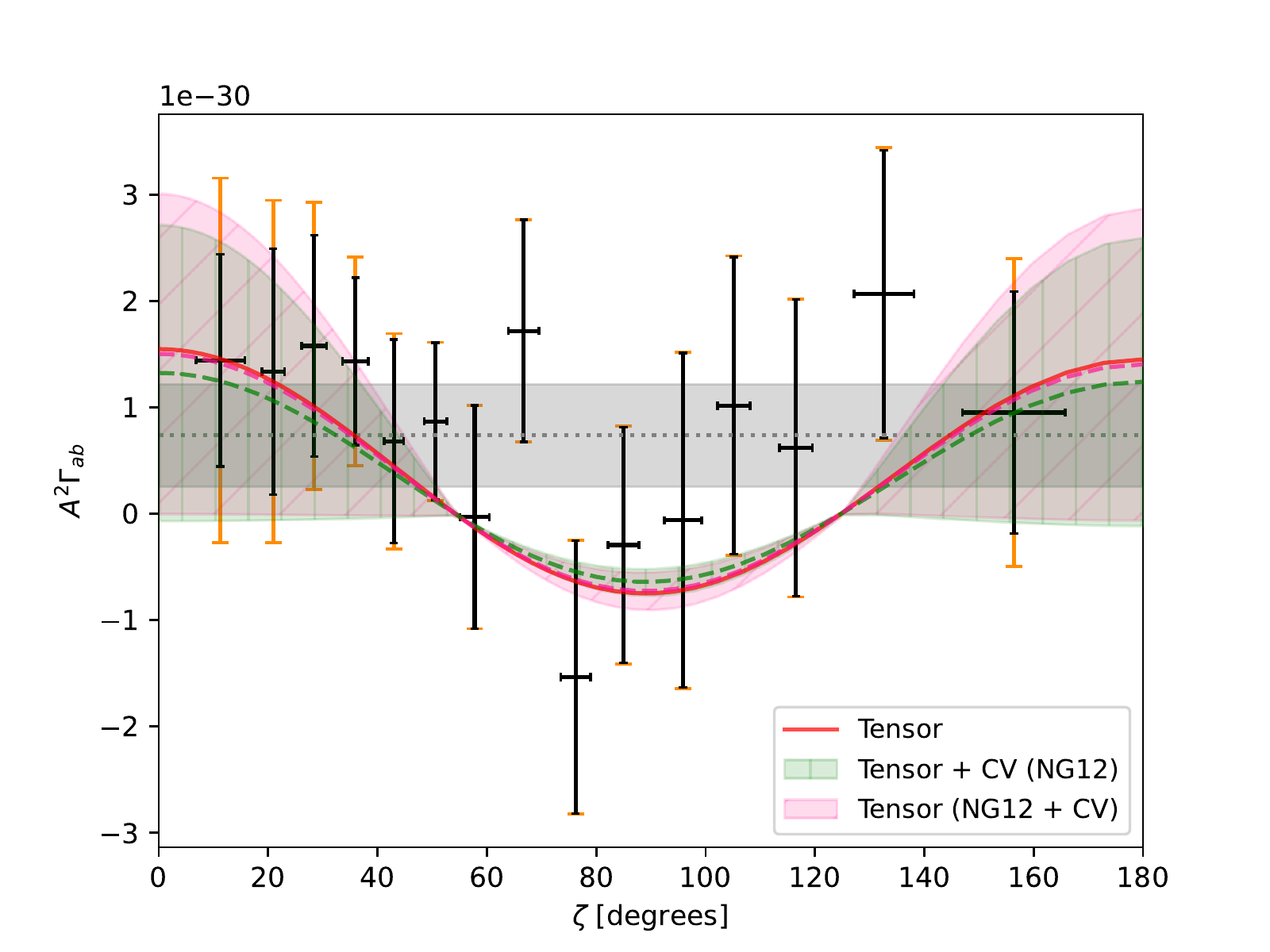}
		}
    \caption{Sampled posteriors and best fit curves for the subluminal tensor correlation with the NANOGrav 12.5 year cross correlation measurements (black bars, orange ones are effective error bars accounting for the cosmic variance). The best fit {uncorrelated Gaussian random noise (gray, dotted line shows mean, non-hatched horizontal band shows $1\sigma$)} is presented for reference. Colored--hatched bands show $1\sigma$ cosmic variance uncertainty from the mean of the correlation.}
    \label{fig:tensors_post}
\end{figure}

The results are shown in Figure \ref{fig:tensors_post} and Table \ref{tab:tensorstats}.

First off, we find that the parameter posteriors echo the conclusions we had with the HD with mock and real data: that the mean of the correlation is sufficient for their inference. In this case, it can be seen well that the posteriors for the GW strain amplitude squared and the GW speed are consistent regardless of the consideration of the cosmic variance (Figure \ref{fig:tensors_post}(a)). It can also be noticed that the best fit correlation curves exhibit a more dominant quadrupolar feature (Figure \ref{fig:tensors_post}(b)), or rather a more suppressed octupolar component, than the HD. This is reminiscent of subluminal tensor GW--anchored correlations \cite{Qin:2020hfy, Bernardo:2022rif, Bernardo:2022xzl, Liang:2023ary}. However, without the cosmic variance, we have previously found the significance of subluminal GW modes in the correlations to be subpar, if not disfavored, compared with the monopole \cite{Bernardo:2023mxc}. This is also reflected in Table \ref{tab:tensorstats} where neither of the performance measures ($\overline{\chi}^2$ {and $\log Z$}) come to support subluminal GWs.

On the other hand, an enhancement in the significance of subluminal GWs in the data can be found when the cosmic variance is considered in the sampling. This appears for both ways we take the theoretical uncertainty as part of the analysis. It is worth highlighting that the subluminal GW--anchored correlations with the cosmic variance turn out with a better goodness of fit compared with the monopole and the HD.

\begin{table}[h!]
    \centering
    \caption{Best fit statistics for subluminal tensor modes cross correlations constrained by PTA \cite{NANOGrav:2020bcs}. The performance statistics {presented are the reduced chi-squared, $\overline{\chi}^2$, and the Bayesian evidence, expressed as $\log Z$, e.g., smaller $\overline{\chi}$ mean a better point estimate, and more positive $\log Z$ point to the better model. We additionally present the marginalized statistics for the monopole and an uncorrelated (Gaussian random) noise for comparison purposes}.}
    \begin{tabular}{|r|r|r|r|r|r|} \hline
    \phantom{$\dfrac{1}{1}$} model (data set) \phantom{$\dfrac{1}{1}$} & \phantom{111} $v$ \phantom{111} & $A^2$ [$\times 10^{-30}$] & {$\overline{\chi}^2$} & {$\log Z$} \\ \hline \hline 
    \phantom{$\dfrac{1}{1}$} T (NG12) \phantom{$\dfrac{1}{1}$} & $0.46^{+0.18}_{-0.42}$ & $7.0^{+2.5}_{-2.8}$ & {$0.63$} & {$-6.56 \pm 0.06$} \\ \hline 
    \rowcolor{Gray}
    \phantom{$\dfrac{1}{1}$} \textbf{T + CV (NG12)} \phantom{$\dfrac{1}{1}$} & $\mathbf{0.45^{+0.17}_{-0.44}}$ & $\mathbf{6.0^{+2.1}_{-3.0}}$ & {$\mathbf{0.46}$} & {$\mathbf{-7.30 \pm 0.11}$} \\ \hline
    \rowcolor{LightCyan}
    \phantom{$\dfrac{1}{1}$} \textbf{T (NG12 + CV)} \phantom{$\dfrac{1}{1}$} & $\mathbf{0.46^{+0.19}_{-0.42}}$ & $\mathbf{6.8^{+2.8}_{-3.2}}$ & {$\mathbf{0.56}$} & {$\mathbf{-6.31 \pm 0.08}$} \\ \hline
    \phantom{$\dfrac{1}{1}$} mon \phantom{$\dfrac{1}{1}$} & ------- & $1.80 \pm 0.55$ & {$0.58$} & {$-6.40 \pm 0.10$} \\ \hline
    \phantom{$\dfrac{1}{1}$} {${\cal N}\left(\sigma < 0.34\right)$ (mock)} \phantom{$\dfrac{1}{1}$} & ------- & {$1.47 \pm 0.63$} & {$0.30$} & {$-8.47 \pm 0.41$} \\ \hline 
    \end{tabular}
    \label{tab:tensorstats}
\end{table}

The larger enhancement {in $\overline{\chi}^2$} comes from giving the correlation a random nature rather than it being a fixed curve as it is usually treated. The similar enhancement in the likelihood of the SGWB correlations model was of course found earlier with the HD, and we expect this to be a general result for GW modes regardless of their nature. {As we previously alluded to, the reduced chi-squared is a point estimate where the likelihood reaches its maximum value and may not be representative of the sampled space. Figure \ref{fig:tensors_post}(a) shows the spread in the $\chi^2$ values, which is clearly widest when the cosmic variance is used to unstiffen the tensorial correlations, leading to a better estimate pointwise but overall a weaker Bayesian evidence. Of course, we have seen the same effect with the HD and so to a degree this is not too surprising. Similarly, the canonical use of the cosmic variance as effective noise appears more reassuring, now that it has improved the Bayes factor of the subluminal tensorial correlation against the null hypothesis (uncorrelated Gaussian random noise) by $Z_{\rm HD}/Z_{\rm null} \sim 6-9$. Understandably, this remains a weak evidence against the null hypothesis, but it is the enhancement in the significance by the cosmic variance that we want to associate credit to. We have seen it for the HD and now also for subluminal tensors.}

The main takeaway is clear: that the chances of detecting a stochastic signal (SGWB) are better when considering not just one but two of its moments, the mean and the variance. This is an exciting result, considering that the detection of the nanohertz GWs in the form of the SGWB can be expected to be the next big astronomical breakthrough since the discovery of GWs by ground--based detectors. Furthermore, this observation of subluminal GWs in the nanohertz GW band is tangled with the fundamental question of the existence of a massive graviton \cite{Bernardo:2023mxc}. In addition, constraints on the GW speed give heavy repercussions to the space of viable alternative theories of gravity, as the spectacular GW170817, and its optical counterpart GRB170817, did a few years ago. Constraining the GW speed in the nanohertz GW band that the PTAs are sensitive to will without doubt come with a comparable impact on fundamental physics, recognizing that the SGWB is a playground for a plethora of exotic high energy phenomena. The cosmic variance might just play a key role in this direction in PTA cosmology.

\section{Discussion}
\label{sec:discussion}

Throughout this work, with mock and authentic data, we have examined the impact of theoretical uncertainty in the form of the `cosmic variance' for SGWB analysis with PTA spatial correlations, and presented quite interesting (and definitely intriguing) results in light of the significance of SGWB correlations. To put it briefly, we have shown that the cosmic variance improves the perceptibility of the SGWB correlations in noisy data. Another way of saying this is that the chances of finding a stochastic signal in noise is better when not only one but two of its leading moments are utilized. 

{
Gently put, the likelihoods can get to higher values however the CV is considered, as reflected in Figs. \ref{fig:hd_mock_post}--\ref{fig:tensors_post}. However, using the CV as HD + CV (data) does acquire a lower Bayesian evidence since most of the sampled points come with lower likelihoods. This way of using the CV is nonetheless exploratory as the established way of using it like in the CMB is HD (data + CV) where the significance increases whether we measure it pointwise or with the entire sampled space. This increase in the likelihood and the evidence also appeared for subluminal GW correlation (data + CV), which suggests that the cosmic variance improves the significance of the GW correlations. This is the message of our paper and overall what we meant by saying that the cosmic variance helps out in looking for GW correlations in noisy data. 
}

{We also remark that the small chi-squared values presented in Tables \ref{tab:hdmockstats}--\ref{tab:tensorstats} are suggestive of a degree of overfitting. But, this is associated with the large uncertainty of the correlations data. It is interesting to see if this continues to play out with the more stringent updated data sets.}

Our analysis naturally comes with caveats, particularly, in relation to the use of the cosmic variance, but the message is clear that the theoretical uncertainties help out, if not already a convincing major player, in the hunt for the SGWB. The cosmic variance however might be a too naive interpretation of the SGWB's correlation uncertainty as it assumes innumerable pulsars in the sky, considering there can also arise statistical challenges from having a limited number of pulsars \cite{Johnson:2022uxn}. A more grounded version of theoretical uncertainty of the SGWB correlations have been derived and discussed in detail in \cite{Allen:2022dzg, Allen:2022ksj}, which holds for a finite number of pulsars in arbitrary distributions in the sky, but is limited to the HD correlation. Nonetheless, it is also worth mentioning that the cosmic variance is a measure of the `Gaussianity' of the stochastic field that makes up the SGWB. Future PTA data sets that precisely measure the SGWB may be used to constrain this hypothesis by looking for deviations from Gaussianity.

We note that while we have focused on tensor GWs in this paper to show our point, we expect our results to hold generally for GWs excited by, say, scalar or vector gravitational degrees of freedom. We also stress that in contrast with \cite{Bernardo:2023mxc}, which excludes theoretical uncertainty in the fitting, we in addition do not consider the (uncorrelated) common spectrum process amplitude in the present paper, to focus only on the correlations synonymous with the SGWB \cite{Allen:2023kib}. The takeaway is simply that a stochastic signal is more than just its mean, even if for most purposes this first moment turns out to be sufficient.

To end, we remark that the SGWB looks like it is almost ready to appear on the horizon of PTA data \cite{NANOGrav:2020spf}. The detection of this would without a doubt be a significant milestone in this golden era of GW astronomy. This opens up a window to the exciting and exotic high energy physics phenomena of cosmic strings \cite{Ellis:2020ena, Buchmuller:2021mbb, Hindmarsh:2022awe}, phase transitions \cite{NANOGrav:2021flc, Xue:2021gyq}, primordial black holes \cite{Nakama:2016gzw, Chen:2019xse}, and supermassive black hole binaries \cite{Shannon:2015ect, Liu:2021ytq} to name a few. Knowing more precisely what to find in such an uncharted cosmological territory cannot be anything less than an advantage for new physics, and definitely a step toward understanding gravity.


\acknowledgments

This work was supported in part by the National Science and Technology Council (NSTC) of Taiwan, Republic of China, under Grant No. MOST 111-2112-M-001-065.



\providecommand{\href}[2]{#2}\begingroup\raggedright\endgroup


\end{document}